\documentclass[preprint]{aastex}

\usepackage{natbib}
\usepackage{amsmath}
\usepackage{graphicx}

\numberwithin{equation}{section}

\newcommand{\com}[1]{(\#)}
\def\field{{\mathcal E}}  
\def\eunit{{$\mu V {\rm\ m}^{-1} {\rm\ MHz}^{-1}$}}
\def\eunitm{{\mu V {\rm\ m}^{-1} {\rm\ MHz}^{-1}}}

\shorttitle{RESUN Search}
\shortauthors{Jaeger, Mutel, & Gayley}
{\date{\center{\today}}

\begin{document}

\title{UHE neutrino searches using a Lunar target: \\ First Results from  the RESUN search}
\author{T. R. Jaeger, R. L. Mutel, K. G. Gayley}
\email{theodore-jaeger@uiowa.edu, robert-mutel@uiowa.edu, kenneth-gayley@uiowa.edu}
\affil{Department of Physics and Astronomy, University of Iowa, Iowa City, IA 52242}

\begin{abstract}
During the past decade there have been several attempts to detect cosmogenic ultra high energy (UHE) neutrinos by searching for radio \^Cerenkov bursts resulting from charged impact showers in terrestrial ice or the lunar regolith. So far these radio searches have yielded no detections, but the inferred flux upper limits have started to constrain physical models for UHE neutrino generation. For searches which use the Moon as a target, we summarize the physics of the interaction, properties of the resulting \^Cerenkov radio pulse, detection statistics, effective aperture scaling laws, and derivation of upper limits for isotropic and point source models. We report on initial results from the RESUN search, which uses the Expanded Very Large Array configured in multiple sub-arrays of four antennas at 1.45~GHz pointing along the lunar limb. We detected no pulses of lunar origin during 45 observing hours. This implies upper limits to the differential neutrino flux  $E^2dN/dE < 0.003$ EeV\ km$^{-2}$\ s$^{-1}$\ sr$^{-1}$ and  $<$0.0003 EeV\ km$^{-2}$\ s$^{-1}$ at 90\% confidence level for isotropic and sampled point sources respectively, in the neutrino energy range $10^{21.6}< E(eV)<10^{22.6}$. The isotropic flux limit is comparable to the lowest published upper limits for lunar searches.  The full RESUN search, with an additional 200 hours observing time and an improved data acquisition scheme, will be  be an order of magnitude more  sensitive in the energy range $10^{21}<E(eV)<10^{22}$ than previous lunar-target searches, and will test Z burst models of neutrino generation.

\end{abstract}


\section{Introduction}
Determining the origin and characteristics of extragalactic cosmic rays is of fundamental importance in astrophysics.  Direct detection of high energy cosmic rays from distant sources is difficult due to interaction with cosmic microwave background photons.  Above a limiting energy of $10^{19.5}$ eV, cosmic rays interact with CMB photons within a few tens of Mpc, producing a high-energy pion shower \citep{Greisen:1966}.  The pions quickly decay, producing ultra-high energy (UHE, $E>$ 1~EeV = $10^{18}$~eV) neutrinos that can propagate unimpeded directly to the Earth. This limit, known as the Greisen-Zatsepin-Kuzmin (GZK) limit,  predicts that UHE particle processes thought to occur in distant AGN cores and hypernovae cannot be seen directly at Earth, but must be traced via the secondary production of UHE neutrinos.  Indeed, recent measurements of the cosmic ray spectrum at UHE energies \citep{Yamamoto:2008} show a clear drop above the GZK limit, consistent with the GZK prediction. Other, more speculative, sources for UHE neutrinos include decaying super-massive particles, such as magnetic monopoles, and cosmic topological defects \citep{Stanev:2004}. Although not yet detected, these UHE neutrinos will provide an fundamentally new window for studying the physics of AGN and possibly exotic astrophysical particles.

% Figure 1 - Cosmic Ray Spectrum
\begin{figure}[h]
\begin{center}
\includegraphics[width=6.25in]{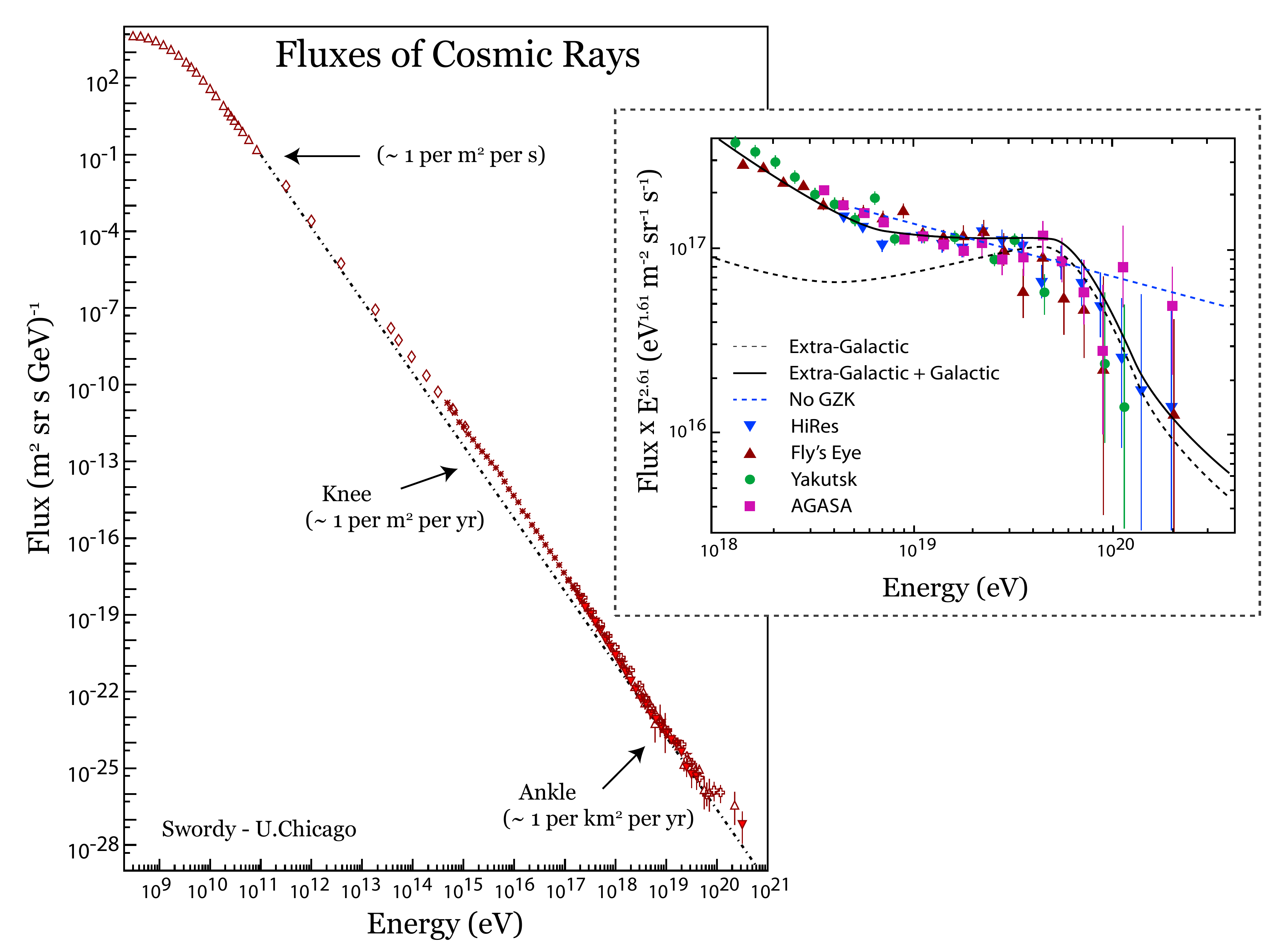}
\caption{Observed Cosmic Ray spectrum across a wide range of energies \citep{Cronin:1997}. Inset shows a detailed view of the "Ankle" region located at the Ultra High Energy end of the spectrum, from \citet{Waxman:2009}. Recent measurements at UHE energies have revealed a drastic attenuation of the CR spectrum at $10^{19.6}$~eV, consistent with the predicted GZK cutoff. The ordinate of the inset has been scaled by a factor of $E^{2.61}$ to better illustrate the CR flux drop-off.}
\label{Fig:CRSpecGZK}
\end{center}
\end{figure}

\begin{figure}[h]
\begin{center}
\includegraphics[width=4in]{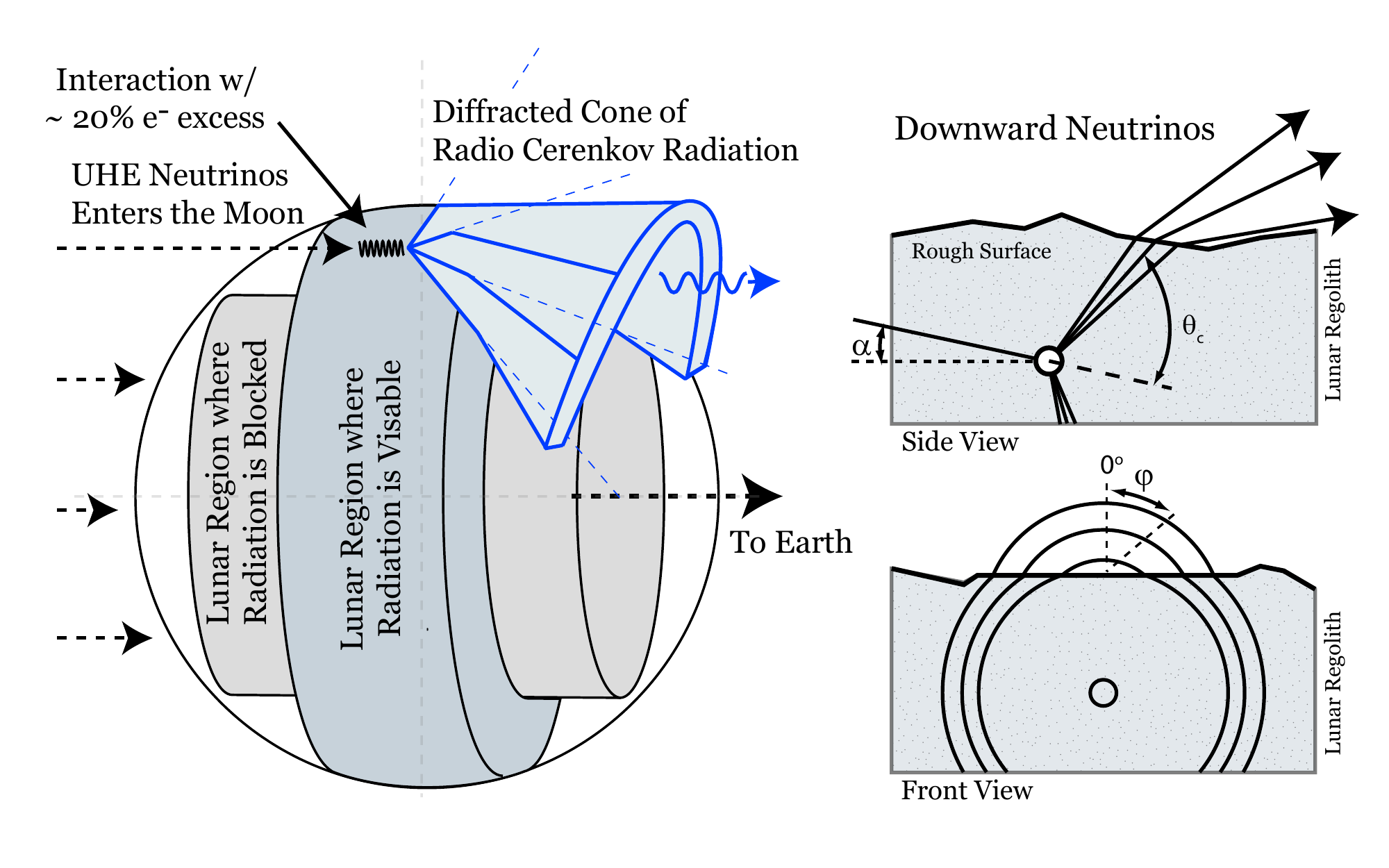}
\caption{UHE Neutrino interaction with lunar regolith matter. A neutrino-parton interaction results in a hadronic shower. The shower creates an energy cascade of $e^+e^-$ pairs with a 20\%  charge excess and consequent \^Cerenkov radiation cone. The cone partially escapes the lunar surface, where an observer sees a nanosecond-duration radio pulse.}
\label{Fig:Interaction1}
\end{center}
\end{figure}

Since the interaction cross-section of neutrinos with matter is notoriously small, large targets are required to detect UHE neutrinos.  A promising approach is the observation of radio bursts resulting from UHE neutrino interactions with the Moon (Fig.~\ref{Fig:Interaction1}).  \citet{Askaryan:1962} calculated that UHE particles would interact with bulk materials such as ice, salts, or the lunar regolith.  The regolith interaction produces a shower of particles with a resulting a 20\% electron excess.  As the electrons are moving at the free-space speed of the incident neutrino $\sim$ c, larger than the phase speed of light in the medium, they produce a very short duration (hence wide bandwidth) burst of radio \^Cerenkov radiation. The radiation is concentrated in a thin cone visible by cm-wavelength radio telescopes.  Askar'yan's prediction and many details of the particle interaction have been directly verified with pulsed electron bunches and salt and ice targets at the Stanford Linear Accelerator \citep{Saltzberg:2001,Miocinovic:2006,Gorham:2007} . The pulse emission was shown to be broadband, linearly polarized, with nanosecond characteristic timescales, and corresponding bandwidth 1-10 GHz, depending on the observer's location in the radiation cone.

% Figure 2 - Neutrino Interaction

Searching for radio bursts with such small time scales is problematic.  Although the peak flux density is predicted to be very intense, e.g. of order $10^5$ Jy at 1~GHz \citep{Alvarez-Muniz:2000}, interference and radiometer noise statistical fluctuations can mimic the desired signals.  Hence, well-designed detection schemes must strongly discriminate signals of lunar origin from 'accidental' (statistical fluctuation) pulses.  The first pioneering search was performed in 1995 when \citet{Hankins:1996} observed the Moon in a band from 1200 -1700 MHz using the Parkes telescope in Australia. They searched for pulses dispersed by the EarthÕs ionosphere, using the ionospheric dispersion delay to discriminate against terrestrial RFI pulses.  A large fraction of their 12 hour experiment was spent pointed at the lunar center, not realizing that only on the lunar limb could \^Cerenkov emission escape in the direction of the observer.  
	
The second lunar radio search, known as GLUE (Goldstone Lunar UHE Neutrino Experiment, \citet{Gorham:2004}) employed the Goldstone 70m and 34m telescopes with a large (22 km) separation to search for delayed pulse coincidence.  They also found no pulses of lunar origin in 120 total hours of observation, but the sampled fraction of the lunar limb was limited by the small primary beams of the antennas compared with the angular size of the Moon. \citet{Beresnyak:2005} described a search using the 64-m diameter Kalyazin telescope using four simultaneous frequency bands between 0.6 and 2.3~GHz with the telescope pointed along the lunar limb. They observed for a total of 31 hours, but detected no pulses with the expected frequency-dependent differential delay from a lunar source. \citet{Scholten:2009a} reported on first results from the 'nuMoon' search using the Westerbork Synthesis Radio Telescope in several bands within the frequency range 115-180~MHz. They had no detections in a preliminary analysis of the first 20 hours of observation, but are continuing to observe. Future lunar radio searches using existing or planned radio telescopes have been discussed by  \citet[][LOFAR]{Scholten:2006},  \citet[][GMRT]{Panda:2007},  \citet[][SKA]{James:2009}, and \citet[][ATCA]{James:2009a}. Finally, we note that several terrestrial searches for \^Cerenkov radio bursts expected from neutrino interactions with the Earth's polar ice sheets have been reported e.g. FORTE \citep{Lehtinen:2004}, RICE \citep{Kravchenko:2006}, ANITA-lite \citep{Barwick:2006} and ANITA \citep{Gorham:2009}. So far, no cosmogenic UHE neutrinos have been detected by any of these experiments.

In this paper we describe the physical and technical considerations relevant to searches for \^Cerenkov radio pulses  resulting from UHE neutrino interactions with the Moon.  Although several previous papers have described various aspects of this scheme, our goal is to fully describe the relevant physical and technical calculations in a self-contained manner. In addition, we apply scaling laws derived from the recent analytic effective aperture calculation of \citet{Gayley:2009} and introduce a new graphical representation of non-detection upper limits which better accounts for the non-detection probability as a function of predicted model flux.

We first discuss the neutrino-regolith interaction physics, properties of the resulting \^Cerenkov radiation, the effective detection aperture, and statistical considerations for multiple antenna detection schemes.  We apply these to derive the sampled neutrino energy range and isotropic flux as a function of  the energy-dependent pulse profile and detector parameters. We then describe the RESUN search (Radio EVLA search for UHE Neutrinos), a radio search using the EVLA\footnote{Expanded Very Large Array, near Socorro NM, operated by the National Radio Astronomy Observatory.}. We summarize the experimental setup, including the use of multiple sub-arrays and high-speed pulse detection hardware to optimize sensitivity to short-timescale pulses. Finally, we report initial results of the search, and compare them with previous radio searches. 

\section{Physics of the UHE Neutrino Lunar Interactions}
\label{sec:detection}

\subsection{Neutrino-Regolith Interaction}
\label{sec:properties-interaction}
Incident UHE Neutrinos interact with baryons in the lunar regolith with an energy-dependent cross-section  \citep{Reno:2005}
\begin{equation}
\sigma(E) = 1.57\times10^{-31}{\left( \frac{E}{ZeV} \right)}^{1/3}\ \ {\rm cm}^2.
\end{equation}
where ZeV = $10^{21}$ eV. The corresponding mean free path of the neutrino can be expressed as
\begin{equation}
\label{eqn:mfp-nu}
\lambda_{\nu}(E)={\left[\frac{m_H}{\sigma(E)\rho}\right]}= 35.7\mathrm{\ km\ } {\left( \frac{ZeV}{E} \right)}^{1/3}
%\lambda_{\nu}(E)=77 \; \mathrm{km} \cdot(\frac{E_0}{E})^{1/3}
\end{equation}
where $E$ is the neutrino energy and $\rho=1.8 \mathrm{\ gm\ cm}^{-3}$ is the mass density of baryons in the regolith \citep{James:2009}.  The neutrino does not impact baryons {\it per se}, but instead with their interior, point-like constituent partons (quarks and/or gluons).  The interaction results in a hadronic shower with approximately 20\% of the initial neutrino energy.  This shower quickly evolves into a short-lived current source via electron entrainment and positron annihilation \citep{Alvarez-Muniz:2008}. This transient current creates a pulse of coherent \^Cerenkov radiation, since the electrons' speed ($c$) is greater than the local phase speed of light in the medium.  

The pulse is emitted in a cone of half-angle 
$$\theta_c = \mathrm{acos}(\frac{1}{n_r})\sim55^{\circ}$$
where $n_r = 1.73$ is the refractive index of lunar regolith \citep{James:2009}.  The pulse is broadband with a characteristic timescale $\tau = c/L_s$, where the shower  length $L_s$ in regolith is given by \citep{Scholten:2006}
\begin{equation}
\label{eqn:len-shower}
L_s(E)=1.64+0.08\log\left(\frac{E}{ZeV}\right)\mathrm{\ m}
\end{equation}          
The corresponding shower timescale is $\tau \sim 5$ nsec with a small dependence on energy. 
The remaining $80\%$ of the neutrino energy goes into one of two different reactions.  For neutral-current (NC, branching ratio $\sim$ 1/3) interactions, the neutrino is scattered and retains the remaining energy.  In the charge-current (CC, branching ratio $\sim$ 2/3) case, a muon, electron, or tau lepton is produced with the remaining $\sim 80\%$ of the energy.  While muons and tauons are absorbed by the surrounding medium, electrons will initiate a electromagnetic cascade which in turn creates a secondary Cerenkov cone.  However, the EM cascade is considerably elongated by the Landau-Pomeranchuk-Migdal (LPM) effect \citep{Alvarez-Muniz:2000}, forming a narrow emission cone which has a negligible contribution to the total emission at most observing angles.

% Figure 3 - interaction detail
\begin{figure*}[ht]
\begin{center}
\includegraphics[width=4in]{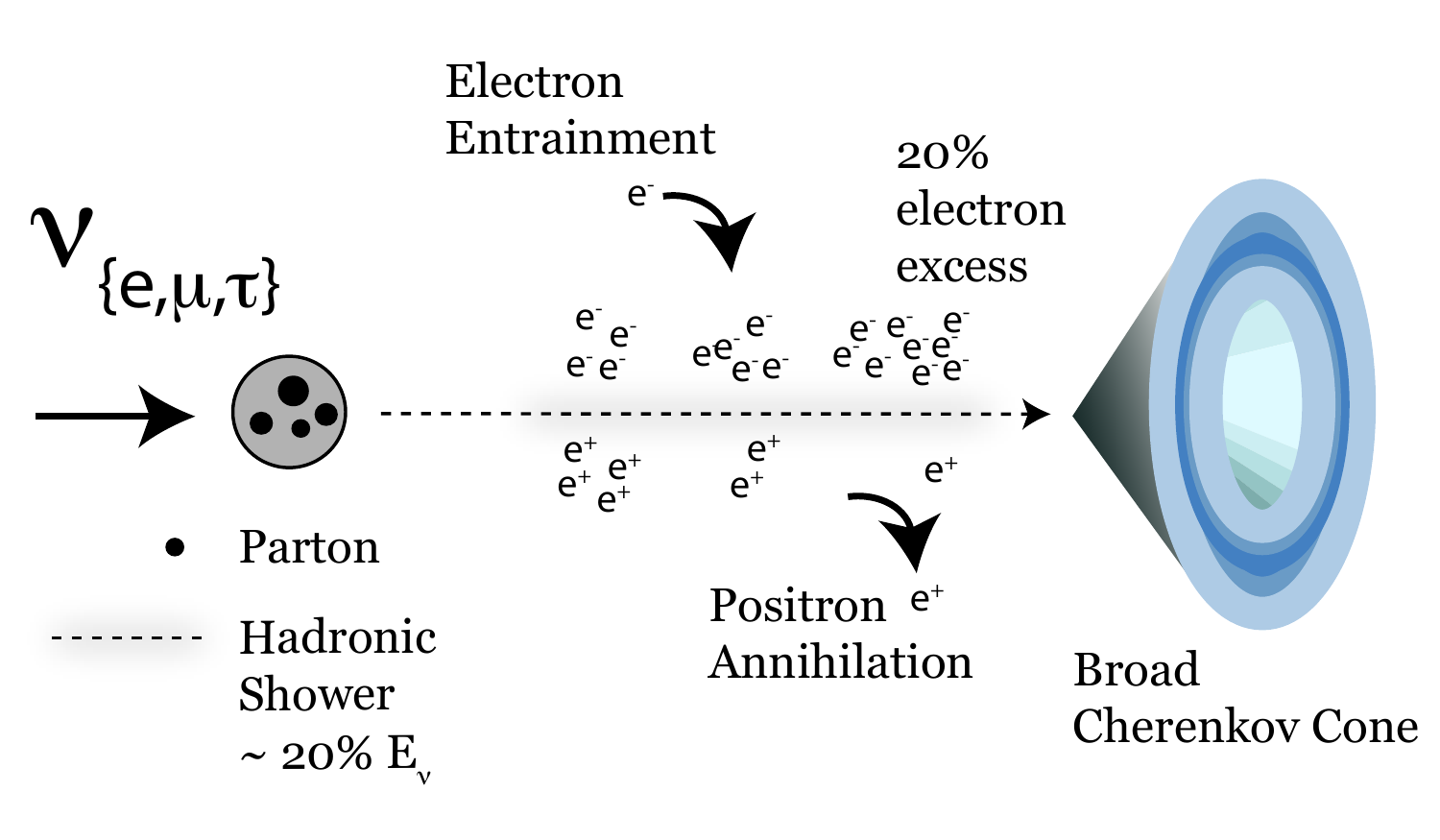}
\caption{Sketch of the neutrino-parton interaction. The hadronic shower carries 20\% of the neutrino energy, eventually in $e^+e^-$ pairs. The positrons are lost more quickly, resulting in a net 20\% $e^{-}$ charge excess. The charged current emits a \^Cerenkov pulse with a cone width given by equation \ref{eqn:cone-width}.}
\label{Fig:interaction-detail}
\end{center}
\end{figure*}

\subsection{Properties of the Cerenkov Radiation Pulse}
\label{sec:properties-cerenkov}
The hadronic shower creates a \^Cerenkov radiation cone whose peak amplitude and angular width depend on energy $E$ and observation frequency $\nu$.  The  cone's full angular width at $e^{-1}$ in lunar regolith  can be written \citep{James:2009}
\begin{eqnarray}
\label{eqn:cone-width}
\Delta\theta_H(E,\nu)= 
2.9^{\circ}  \left[\frac{\mathrm{GHz}}{\nu}\right] {\left[{1.15+0.075\log\left(\frac{0.2E}{ZeV}\right)}\right]}^{-1}
\end{eqnarray}
where we have multiplied the \citet{James:2009} expression by 1.2 to convert from half-wifth to $e^{-1}$ width. At $E=1$ ZeV, the resulting cone widths vary from $17.6^{\circ}$ at 150~MHz to $ 1.3^{\circ}$ at 2~GHz.    

In the absence of regolith attenuation, the peak electric field strength ($\theta =\theta_c$) of a lunar Cerenkov pulse received at the Earth can be parameterized as \citep{Gorham:2000,James:2009}
\begin{equation}
\field_{max}^c(E,\nu) =0.51  \left[\frac{0.2E}{ZeV}\right] \left[\frac{\nu}{\nu_0}\right] {\left[1+\left(\frac{\nu}{\nu_0}\right)^{\alpha}\right]}^{-1}\ \ \mu{\rm V}\ {\rm\ m}^{-1}\ {\rm MHz}^{-1}
\label{eqn:emax0}
\end{equation}
where $\nu_0=2.32$ GHz, $\alpha=1.23$ for regolith material \citep{James:2009} and $\nu$ is the observation frequency.  
For Phase-A of the RESUN search  ($\nu_0$=1.45~GHz,  $\Delta\nu=$ 50~MHz), the peak electric field of the \^Cerenkov pulse at the telescope, including transmission  loss the lunar surface, is
\begin{equation}
\field_{max}^c(E)= 0.041  \left(\frac{E}{ZeV}\right) \ \eunitm
\label{eqn:emax-resun}
\end{equation}
and the $e^{-1}$ width (ignoring the small effect of the sin terms in \ref{eqn:emax}) of the \^Cerenkov cone is $\Delta\theta_H\sim2.1^{\circ} $ for all $E\geq E_0$.

Unfortunately, the burst radiation will be strongly attenuated by the local medium as it propagates to the surface from the shower interaction in the interior. The $e^{-1}$ attenuation length for the electric field in lunar regolith is \citep{Olhoeft:1975}
\begin{equation}
\label{eqn:radio-depth}
\lambda_{\field}=18\; \mathrm{m} \cdot \left[\frac{GHz}{\nu}\right]
\end{equation}
Hence, the electric field strength from each escaping ray path along the \^Cerenkov cone must be corrected by a factor 
\begin{equation}
\field(z) = \field(0)\int_0^{z\sec(\theta)} e^{-\left({s}/{\lambda_{\field}}\right)}\,ds\ ,
\end{equation} 
where $z$ is the vertical depth of the shower and $\theta$ is the ray path angle to the surface. 
The short attenuation length implies only glancing angle impacts produce observable Cerenkov cones.  Neutrinos with large downward angles create radiation cones which are completely absorbed or suffer total internal reflection.  

The angular dependence of the \^Cerenkov cone electric field is \citep{Alvarez-Muniz:2006}
\begin{equation}
%%\label{eqn:emax-theta}
\field^c(\theta,E,\nu)=\field_{max}^c(E,\nu)\ e^{-\zeta(\theta,E,\nu)^2} \ \frac{\sin(\theta)}{\sin(\theta_c)} 
\label{eqn:emax}
\end{equation}
where
$$ \zeta(\theta,E,\nu)=\frac{\theta-\theta_c}{\Delta\theta_H(E,\nu)}$$

Equation \ref{eqn:emax} represents the burst electric field profile interior to the lunar surface.  Of greater interest is the transmitted electric field after propagating through the regolith-vacuum interface. The transmission across this interface is a complicated function of both incident angle and polarization \citep{Williams:2004, Gusev:2006}.  However, \citet{Gayley:2009} argue that the transmission can be approximated by a constant value $t_{\parallel} \approxeq 0.6$ without significant error.  With this in mind, we write the maximum free-space electric field strength as
\begin{equation}
\label{eqn:emax-tx}
\field_{t}^c(E,\nu) = 0.6 \ \field_{max}^c(E,\nu)\ .
\end{equation}

% Signal to noise Ratio
\subsection{Signal-to-Noise Ratio and Minimum Detectable E-Field}
\label{sec:detection-min}
To detect the \^Cerenkov burst, the electric field given by Equation \ref{eqn:emax0} must  exceed the minimum receiver sensitivity for each detector element (single antenna).  Note that this is different from interferometer arrays, in which signals are co-added coherently. The RMS electric field fluctuations at the feed of a radio telescope with effective area $A_e$, system temperature  $T_{sys}$, and bandwidth $\Delta\nu$ can be calculated by equating the power associated with a radiation field having and RMS electric field $\field$ illuminating the telescope area $A_e$ with the shot noise power of the receiver,
\begin{equation}
\nonumber
\frac{{\field}^2}{Z_0} \Delta\nu=\frac{\eta\ k_bT_{sys}}{A_{e}}
\end{equation}
so that the RMS field due to detector noise fluctuations is given by
\begin{equation}
\field^d={\left(\eta\ \frac{k_bT_{sys}Z_0}{A_e\ \Delta\nu}\right)}^{\frac{1}{2}}
\label{eqn:min-antenna-efield}
\end{equation}
where $\eta$ is a dimensionless constant which depends on the polarization properties of the telescope feed and radiation, $k_b$ is Boltzmann's constant, and $Z_0 = 377\ \Omega$ is the impedance of free space.  For a given detector scheme, the minimum detectable electric field is a multiple $N_{\sigma}$ of the RMS field given by the detection threshold as described in section \ref{sec:detection-false}.  For Phase-A of the RESUN search, ($N_{\sigma}$ = 3.98,$\eta=2$ for linearly polarized radiation and circularly polarized feeds, $T_{sys}\sim$ 120\ K pointing on the lunar limb, $A_e = 343$ m$^2$) this evaluates to
\begin{equation}
\field_{min}^t =\ N_{\sigma}\cdot\field^d = 0.034\ \eunitm
\label{eqn:emin-resun}
\end{equation}
By equating equations \ref{eqn:emax-tx} and \ref{eqn:emin-resun}, we find that the minimum detectable incident neutrino energy for RESUN-A is $E_{min}=1.38$\ ZeV.

% Setting the threshold
\subsection{Setting the Threshold: Accidental Trigger Rates}
\label{sec:detection-false}
At each detector, a threshold level must be set to discriminate against false pulse detection arising from statistical noise fluctuations. Assuming the receiver noise voltage obeys Gaussian statistics and that the receiver maintains a linear response within the voltage range of the fluctuations (see section 4), the  probability of a sample signal $\sigma$ exceeding a specified threshold $\sigma_{t}$ is given by: 
\begin{equation}
\label{eqn:prob-single}
p_o(\sigma > \sigma_{t}) = \frac{1}{\sqrt{2\pi}}\int_{\sigma_{t}}^{\infty}
exp{\left(-\frac{{\sigma}^2}{2}\right)}d\sigma = {\rm erfc}\left(\frac{\sigma_{th}}{\sqrt{2}}\right)
\end{equation} 
where $erfc$ is the complementary error function.  With fast sample times, the false signal rate can be quite high, even for very high threshold values. For example, with a 100 MHz sample rate (10 ns samples) and a $6\sigma$ threshold, one expects an `accidental'  trigger approximately once every 5 seconds.        

% Figure 4 - Accidental rate vs sigma
\begin{figure*}[b!]
\begin{center}
\includegraphics[width=4in]{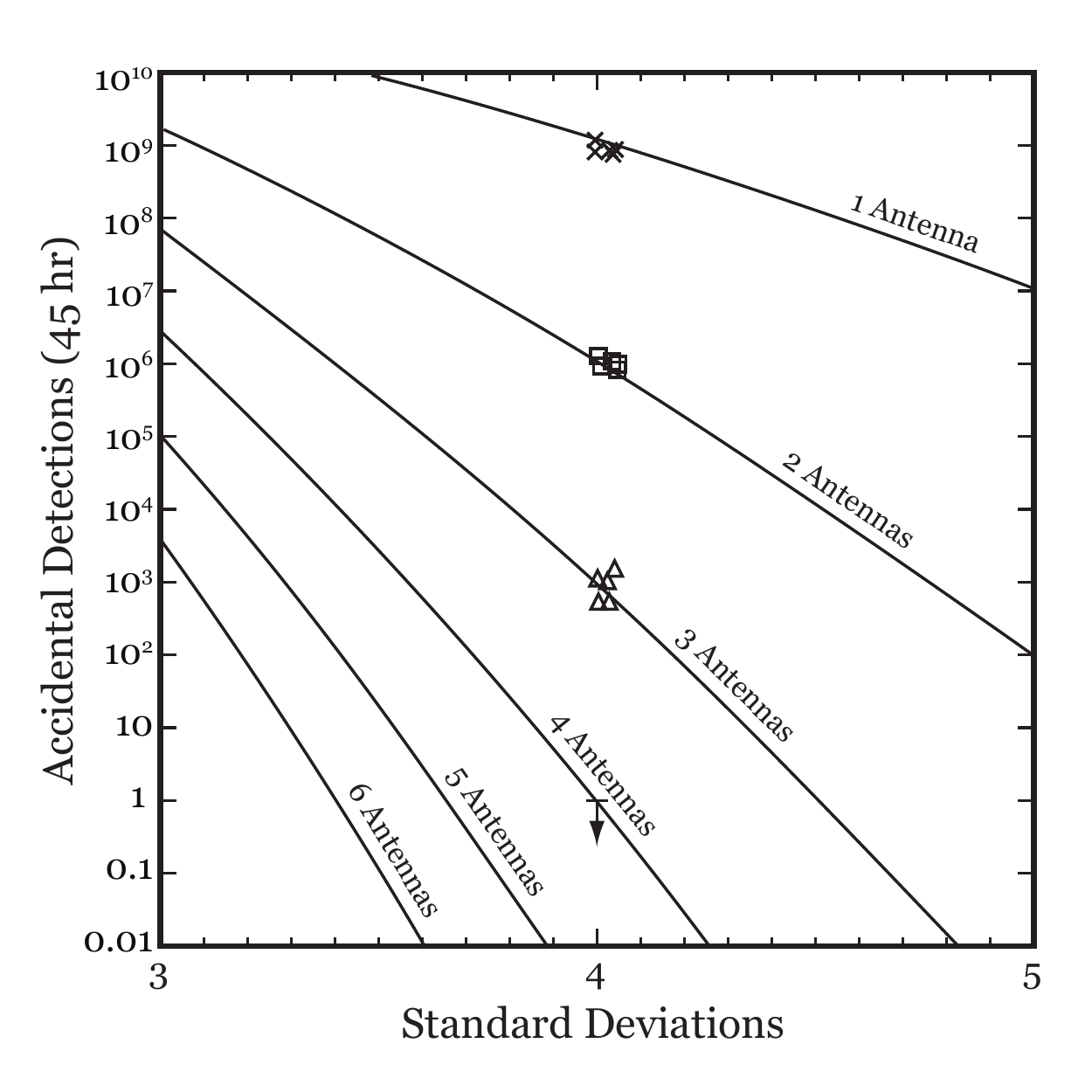}
\caption{Expected number of accidental detections per second versus threshold level calculated using equation \ref{eqn:accid-prob}. Accidental rates are shown for arrays of one to six telescopes using 10 ns sampling time and 140 ns. delay tolerance window. The measured accidental detection rate for five sample one-minute intervals (scaled to 50 hours) for one telescope (x's), two telescopes (squares), three telescopes (triangles), and four telescope (upper limit) detection schemes are shown using a 4$\sigma$ threshold are in excellent agreement with the noise model. }
\label{Fig:Event-Sigma}
\end{center}
\end{figure*}
       
The primary beam of a radio telescope will illuminate an area on the lunar surface with a range of geometrical delays corresponding to a window of $N_s$ time samples. Assuming $p_0 << 1$, The probability $p_w$ of a sampled time exceeding the threshold level in the window of $N_s$ samples is simply 
\begin{equation}
p_w(\sigma_{t},N_s)=1-[1-p_o(\sigma_{t})]^{N_s}\sim N_s\cdot p_o(\sigma_{t})
\end{equation}
For an array of $N_a$ telescopes whose primary beams are coincident on the Moon, the joint probability $p_j$ that a statistical fluctuation will exceed the threshold level at all telescopes is the product of the individual antenna detection probabilities, where in general, each antenna may have a different signal to noise ratio and corresponding threshold level.  Thus, the joint probability of an accidental trigger in time window of $N_s$ samples (after correction for geometrical delays) for $N_a$ antennas is:
\begin{equation}
 p_{j}(\sigma_{t},N_s,N_a)= N_s^{\left(N_{a}-1\right)}\prod_{i=1}^{N_{a}}{p_0({\sigma_t}_i)}
 \label{eqn:accid-prob}
\end{equation}
where we have accounted for the possibility of different threshold levels ${\sigma_t}_i$ for each antenna.
The result is a dramatic decrease in the accidental trigger rate, even with small arrays of a few telescopes.  For a given experiment, the threshold level must be set high enough that the probability of an accidental trigger during the observation is highly unlikely. 

%% Expected detection rate
\subsection{Expected detection rate: Effective Aperture Calculation}
\label{sec:properties-aperture}
Calculation of the expected detection rate is a difficult problem because of the non-trivial geometry of the partially escaping radiation cone, attenuation in the lunar regolith, the effects of random surface roughness, the frequency-dependent detector sensitivity, and the 2-dimensional dependence on incident neutrino angle. Several groups have solved this problem for specific experiments by developing Monte Carlo simulations, following ray paths for an variety of neutrino flux models \citep{Gorham:2000, Williams:2004, Gusev:2006, James:2007, James:2009b}. More recently, we  have developed an analytic approach that expresses the expected detection rate  as a function of characteristic physical parameters \citep{Gayley:2009}, which we summarize here. 

The complete system of lunar target and detector can be parameterized by  a direction-dependent detection aperture, which when multiplied by the incident neutrino flux, gives the detection rate in that energy bin and direction. In the following, we consider only isotropic neutrino fluxes, although this approach can be generalized to a direction-dependent case \citep[e.g.,][]{James:2007}.

The effective aperture $A_e$ can be written as the product of the cross-sectional area of the Moon, the total solid angle of incident neutrinos ($4\pi$ for isotropic neutrino flux), and  a detection probability $P$ which takes into consideration the interaction specifics,
\begin{eqnarray}
\label{eqn:Aeff}
A_e(E,\nu,\field_{min}^d)=A_{0} \cdot P(E,\nu,\field_{min}^d)\ ,
\end{eqnarray}
where $E$ is the neutrino energy, $\nu$ is the observing frequency, $\field_{min}^d$ is the minimum detectable electric field strength of the detector (see Section \ref{sec:detection-min}), and $A_{0}$ is the geometric lunar aperture \citep{Williams:2004},
\begin{eqnarray}
%%\label{eqn:aeff-max}
A_{0}=4\pi \cdot (\pi R_m^2) \nonumber \ ,
\end{eqnarray}
where $R_M$ is the radius of the Moon. Note that the aperture has units of physical area times solid angle.
The probability $P$ is the fraction of neutrinos entering the Moon at energy $E$ whose radio pulse will be detectable with a telescope (detector) having minimum electric field sensitivity $\field_{min}^d$ at frequency $\nu$.  
We now summarize the calculation of $P$ detailed in \cite{Gayley:2009}.

An approximate expression for $P$ can be written
\begin{equation}
\label{eqn:detection-probability}
P=P_0 \left(f_0 \Delta\theta_H+ \frac{16}{3\pi^{3/2}} \sigma_s +  \frac{16}{3} \alpha_0\right)\ ,
\end{equation}
where
$$P_0=\left[\frac{n_r^2-1}{8n_r}\right] \frac{\lambda_{\field}}{\lambda_{\nu}} f_0^3 \Delta\theta_H  $$
and the three bracketed terms represent the contributions from downward incident neutrinos on a smooth surface, downward neutrinos on a rough surface, and upward traveling neutrinos respectively.
In the first term, the  dimensionless scaling parameter $f_0$ is the ratio of the thickness of the \^Cerenkov cone at the  $e^{-1}$ full thickness to the thickness at the minimum detectable electric field $\field_{min}^d$ (cf. Fig. 2, \cite{Gayley:2009},
\begin{equation}
\label{eqn:width-emin}
f_0 (E,\nu)=
\sqrt{\ln \left[\frac
    {\field_t^c(E,\nu)}
    {\field_{min}^d(\nu)}
 \right]}\ .
\end{equation}
In the second term,  $\sigma_s$ is the Gaussian half-width RMS surface roughness angle of the lunar surface.  A self-similar (fractal) model provides a good fit to radar surface return measurements, which we can write as a function of observing frequency \citep{Shepard:1995,Gayley:2009},
\begin{equation}
\sigma_o(\lambda) = 
%81\degr\arctan\left(0.29{\nu}^{-0.22}\right) / ,
  \sqrt{2}\ {\rm tan}^{-1}(0.14\ \nu^{0.22})\ \sim 0.2\ {\nu}^{0.22}\ ,
\label{sigma-surface}
\end{equation}
where $\sigma_o$ is in radians and $\nu$ is in GHz.
For the RESUN search ($\nu =1.45$~GHz), we obtain $\sigma_s =12.2^{^\circ}$. 
In the third term, $\alpha_0$ is the maximum acceptance angle for upward incident neutrinos which contribute to the escaping radiation cone, and is given by
\begin{equation}
\alpha_0(E)=\frac{1.4\lambda_{\nu}(E)}{2R_M}=1.8^{\circ}\left({\frac{E}{E_0}}\right)^{\frac{1}{3}}
\end{equation}

% Figure - 5 Aperture vs Emin, obs. freq
\begin{figure*}[h!]
\begin{center}
\includegraphics[width=3.6in]{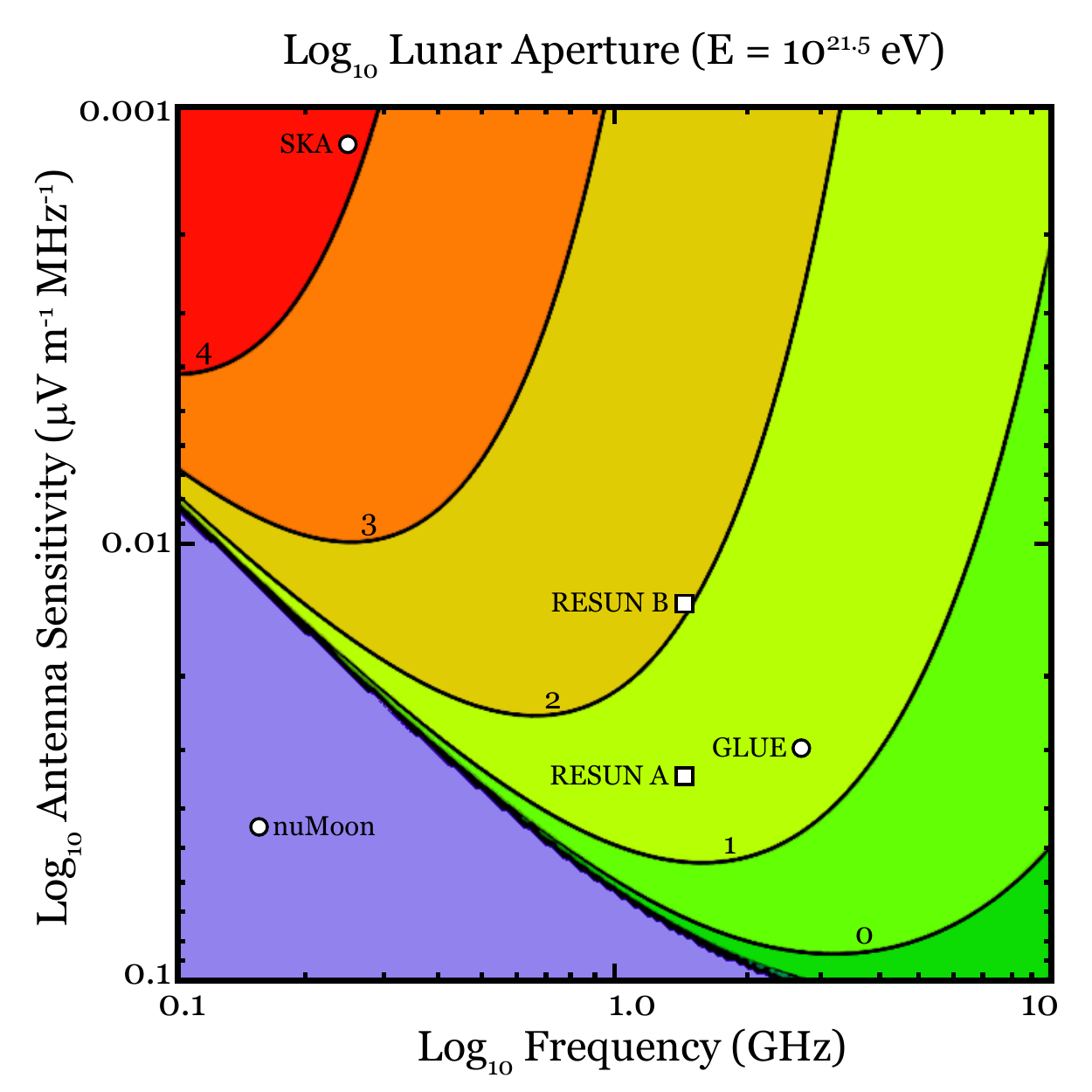}
\caption{Aperture versus detector sensitivity ($\field^t_{min}$) and observing frequency ($\nu$) for a fixed neutrino energy E = $10^{21.5}$~eV. Values for experiments RESUN Phase-A , RESUN Phase-B , GLUE, nuMoon, and SKA are also shown. }
\label{Fig:Aperture-obs-params}
\end{center}
\end{figure*}

% Figure 6 Optimal frequency
\begin{figure*}[ht!]
\begin{center}
\includegraphics[width=3.6in]{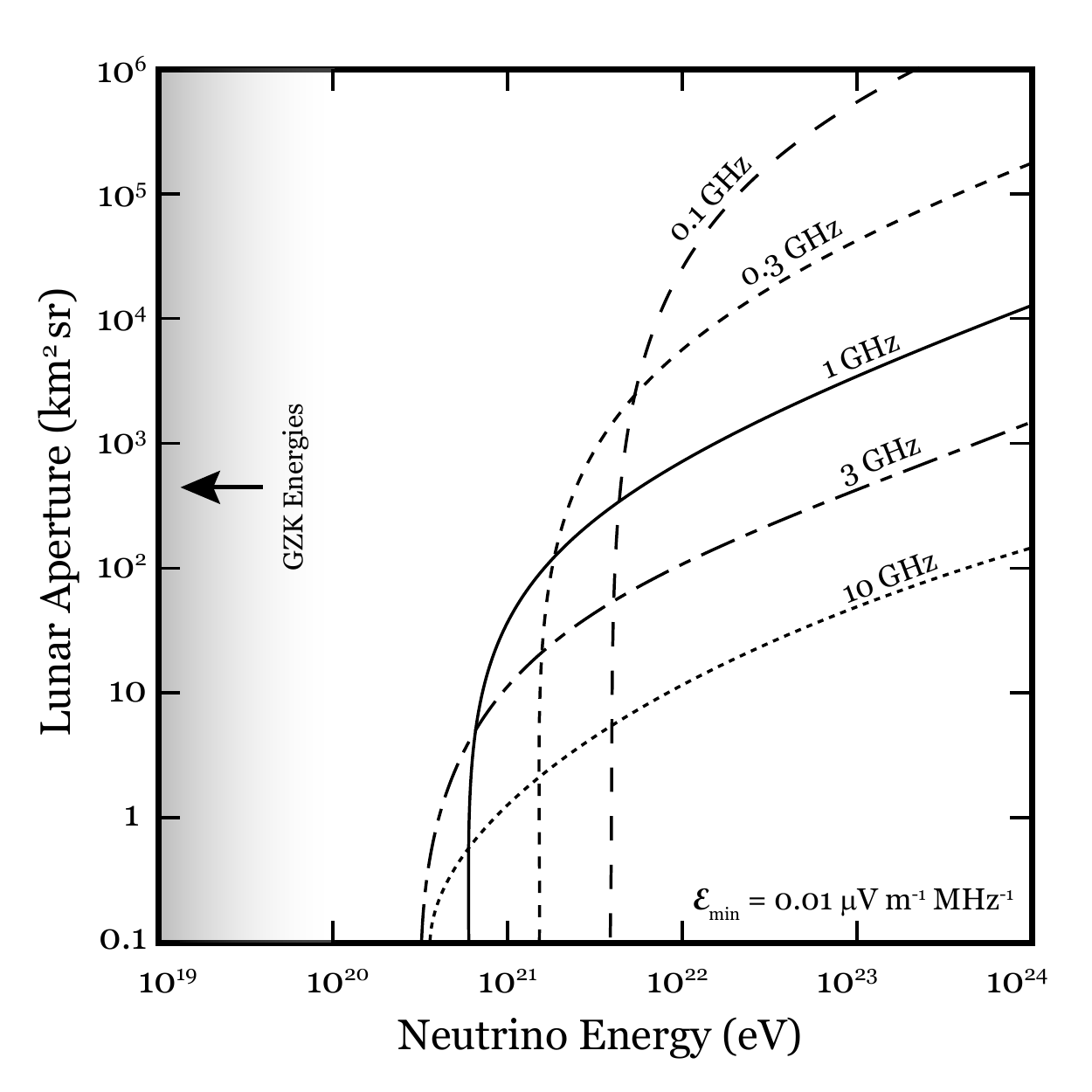}
\caption{Effective aperture versus neutrino energy for fixed detector sensitivity $E_{min}$=\ 0.01 \eunit and observing frequencies between 0.1~GHz and 10~GHz.}
\label{Fig:AEvE}
\end{center}
\end{figure*}

Fig. \ref{Fig:AEvE} shows the effective aperture as a function of neutrino energy for fixed detector sensitivity  ($E_{min}=0.01$ \eunit) and a range of observing frequencies from 0.1~GHz to 10~GHz using equation \ref{eqn:detection-probability}. Note that the sharp low-frequency cut-off is a result of $f_0\rightarrow0$ ,which for fixed telescope sensitivity occurs at higher neutrino energy as the observing frequency is lowered. Frequencies near 1~GHz provide the best compromise between low-energy cutoff and maximum aperture above the cutoff.

\subsection{Flux Limits for Non-Detection}
The expected number of detections $N_{\nu}$ in exposure time $t$ is the product of the neutrino flux and the effective target area integrated over the total observing time, solid angle, and all detectable energies,
%\citep{Lehtinen:2004, James:2007},
\begin{equation}
\label{eqn:event-rate}
N_{\nu}(t,\Omega)=\int_0^t dt^{\prime}\int_{4\pi}d\Omega\int_{E_{min}}^{\infty}dE\ I(E,\hat{n}) \; \hat{A}(E,\hat{n}(t^{\prime})) \; ,
\end{equation}
where $\hat{A}$ is (units of area) is related to the aperture $A_e$ (units of area-sr) by
\begin{equation}
\label{eqn:ae-area}
A_e(E)=\int_{4\pi}\;\hat{A}(E,\hat{n})\;d\Omega
\end{equation}
and $I(E,\hat{n})$ is the differential energy neutrino flux in direction $\hat{n}$.  For the case of an isotropic neutrino flux, we have
$$I(E,\hat{n}) = I(E)$$
where $I(E)$ is the flux per unit steradian. This results in an expected count in time $t$
\begin{equation}
\label{eqn:iso-detect-rate}
N_{iso}=\int_0^t dt^{\prime}\int_{E_{min}}^{\infty}dE\ I(E)\int_{4\pi}d\Omega \; \hat{A}(E,\hat{n})\;,
\end{equation}
or, using eqn. \ref{eqn:ae-area},
\begin{equation}
\label{sec:aeff-limit}
N_{iso}=t\int_{E_{min}}^{\infty}dE\ I(E)\;A_e(E)\;.
\end{equation}
Alternatively, for a point source the neutrino flux is a delta-function in direction,
$$I(E,\hat{n})=I(E)\;\delta(\hat{n}-\hat{n}_s)$$
where $\hat{n}_s$ is the unit normal direction to the source. Substituting into equation \ref{eqn:event-rate}, we have
\begin{equation}
N_{p}=\int_0^t dt^{\prime}\int_{E_{min}}^{\infty}dE\ I(E)\; \hat{A}(E,\hat{n}_s(t^{\prime})\approx t_{p}\int_{E_{min}}^{\infty}dE\ I(E)\; \hat{A}(E)\;,
\end{equation}
where $t_p$ is the total time that the aperture intercepts the source as the Moon moves across the source region.

In order to proceed, we need to consider the energy dependence of the aperture solid angle. The effective aperture directional distribution on the sky  is a complicated calculation which has only been done using Monte Carlo simulations \citep{James:2007}. However, we can use the analytic results of \citet{Gayley:2009} to approximate the solid angle ($\Omega_A$) by noting that at high frequencies ($\nu \gtrsim 1$ GHz) the downward neutrino acceptance angular spread is dominated by the surface roughness angle $\sigma_0$ (eqn. \ref{eqn:detection-probability}) and is much larger than  the upward-directed contribution. Therefore the radial angular spread is approximately the surface roughness angle, which is independent of neutrino energy.    We then approximate $\hat{A}$ as
$$\hat{A}(E) \approx \frac{A_e(E)}{\Omega_A}\ ,$$
where $\Omega_A$ is an equivalent-width estimate of the aperture solid angle, which can be approximated as
$$\Omega_A \approx\; {\sigma_0}^2\ .$$

Collecting terms, we have the expected count rate for a point source 
\begin{equation}
\label{eqn:pt-detect-rate}
N_{p} \approx \frac{t_{p}}{\Omega_A}\int_{E_{min}}^{\infty}dE\ I(E)\; A_e(E)\;,
\end{equation}

 Once the energy-dependent aperture $A_e(E)$ is known, equations  (\ref{eqn:iso-detect-rate}) or (\ref{eqn:pt-detect-rate}) can be differentiated and used to determine the differential neutrino flux ($dN/dE$) as a function of energy. For experiments with a null detection, at 90\% confidence level the neutrino flux upper limit is 2.3 times the differential flux upper limit, assuming Poisson counting statistics. Using to the commonly plotted quantity $F(E)=E^2 dI(E)/dE$,  the upper limit for a non-detection in a given observing time $t$ is 
\begin{equation}
\label{eqn:flux-upperlimit}
F_{iso}(E,t) < 2.3\;\frac{1}{t}\frac{E}{A_{e}(E)}
\end{equation}
while a single point source of neutrinos, the corresponding upper limit is
\begin{equation}
\label{eqn:flux-upperlimit-pt}
F_p(E) < 2.3\;\frac{\Omega_A}{t_p} \;\frac{E}{A_{e}(E)}
\end{equation}
where $t_p$ is the total time that the source is within the aperture solid angle. For lunar search experiments, the beam tracks the Moon's motion on the celestial sphere, so that the mean observation time can be written
$$t_p = 105^{h}\;\times \theta_{t}$$
where $\theta_{t}$ (radians) is the effective angular extent of the aperture solid angle as it transects the point source. If the aperture is dominated by the lunar roughness term in equation \ref{eqn:detection-probability}, we can write
$$\theta_t \approx\;\sqrt{\Omega_A}\approx\;\sigma_0\;.$$

Collecting terms and expressing the frequency dependence explicitly, we can recast eqn. (\ref{eqn:flux-upperlimit-pt}) using approximate expressions above for $t_p$, $\sigma_o$, and $\theta_{eq}$ to obtain the flux detection limit (90\% confidence) for point sources
\begin{equation}
\label{flux-pt-upper-limit-numerical}
F_p(E,\nu)\ \sim 1.2\times10^{-6}\ {\left[\frac{\nu}{GHz}\right]}^{0.22}\frac{E}{A_e(E)}
\end{equation}

% RESUN SEARCH section
\section{RESUN Search }
The RESUN search uses the National Radio Astronomy Observatory's Very Large Array (VLA) to search for \^Cerenkov radio bursts originating from the Moon. The search comprises two phases: RESUN-A, a 45-hour search completed in February 2008, and RESUN-B, a 200-hour search with significantly enhanced limb coverage, wider total bandwidth, and lower pulse detection threshold.  This section describes the experimental design and effective apertures for both searches and reports on results for the RESUN-A search. RESUN-B is scheduled for completion in early 2010.
\label{sec:RESUN-tests}

%%  ARRAY CONFIGURATION
\subsection{Array and receiver configuration}
\begin{figure*}[h!]
\begin{center}
\includegraphics[width=4in]{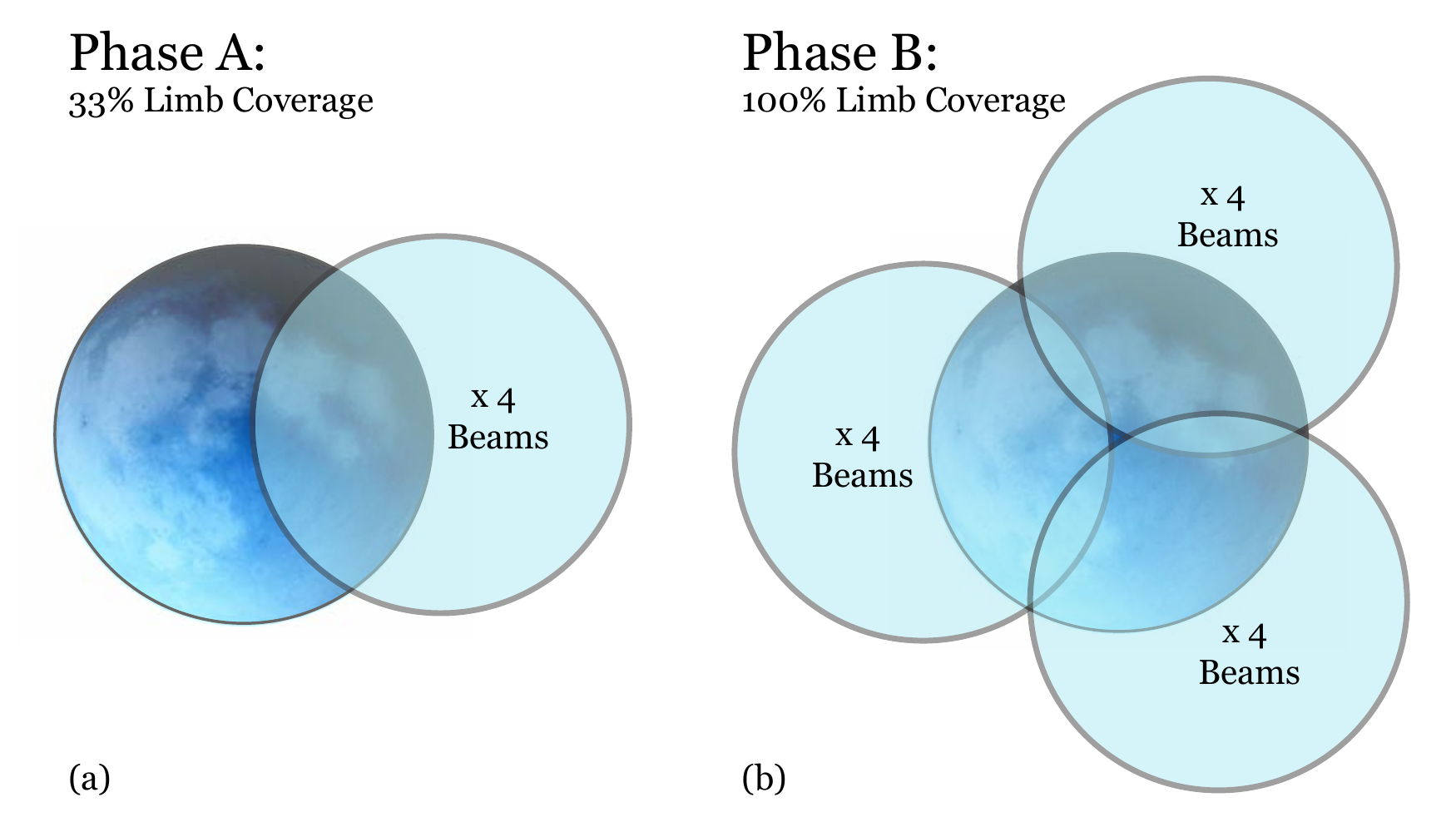}
\caption{Subarray beam geometries for RESUN-A (left) and RESUN-B  (right) experiments.}
\label{Fig:BeamGeometry}
\end{center}
\end{figure*}

The VLA consists of 27 25-m diameter parabolic reflectors with Cassegrain optics in an array whose maximum baseline varies from 1 km to 30 km depending on array configuration. For RESUN-A we used one sub-array of four antennas with antenna separations ranging from 0.6 km to 5.3 km. We observed right-circularly polarized radiation at 1.45 GHz center frequency with a 50~MHz bandwidth. Each antenna in the subarray was pointed to a target on the east limb of the Moon  (Fig.\ref{Fig:BeamGeometry}a). The VLA antenna FWHM primary beam at 1.45~GHz is 30 arcmin. Approximately one third of the lunar limb was illuminated in this configuration. Note that each antenna is an independent pulse detector so the array is not being used in the usual interferometer mode. 

% Figure 7 - Beam geometry

\begin{figure*}[h!]
\begin{center}
\includegraphics[width=5.75in]{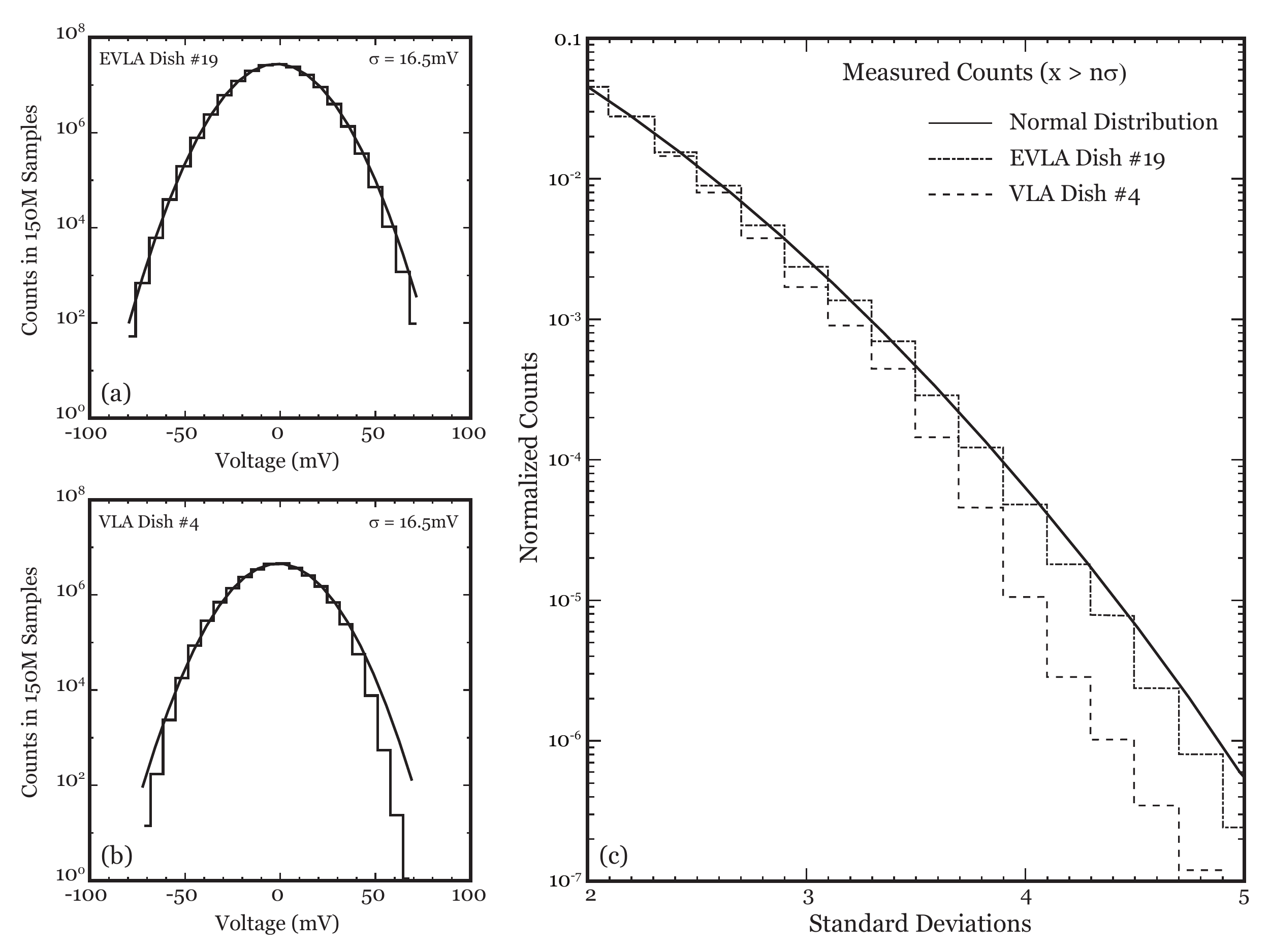}
\caption{Comparison of measured count rates for sample VLA (antenna 4, dashed line), EVLA (antenna 19, dashed-dot line) receivers, and Gaussian statistical model (solid line) versus threshold level for sampled voltage output of square-law detector (proportional to input power).  Note that while the VLA receiver shows significant saturation above $3\sigma$,  there is very little saturation for the ELVA system for signals $\lesssim5\sigma$, well above the RESUN threshold. }
\label{Fig:Good_Ant}
\end{center}
\end{figure*}

The VLA receivers and correlator are currently being upgraded to improve sensitivity and frequency agility \citep{Butler:2006}. The RESUN-A search was scheduled during the transition from the 'classic' VLA system to the upgraded EVLA system. In this hybrid transition mode, both VLA and EVLA antennas were available.  In order to determine the most appropriate system for RESUN, we measured the receiver output voltage dynamic range for both VLA and EVLA receiver systems by comparing histograms of sampled receiver output voltage to a Gaussian statical model (Fig. \ref{Fig:Good_Ant}). A large voltage dynamic range is critical to \^Cerenkov pulse detection experiments, since the calculated pulse amplitudes much larger than typical astronomical sources. The exact threshold value depends on the experimental setup, as discussed in section \ref{sec:detection-false}. We found that the 'classic' VLA receivers were unusable for the RESUN search since they saturated well below the RESUN threshold of $4\sigma$, while the new EVLA receiver systems were in excellent agreement with Gaussian statistics until voltage excursions approached $5\sigma$. Hence we used a subarray comprised entirely of EVLA antennas.  

% Figure 8 - VLA vs EVLA histograms

%% PULSE DETECTION HARDWARE
\subsection{Pulse detection scheme}
\label{sec:RESUN-hardware}

The down-converted waveform at each antenna was sampled and processed using FPGA-based digital signal processors 
developed by the CASPER laboratory at U. C. Berkeley  \citep{Parsons:2008}. Each subsystem module consists of two dual channel 8-bit analog to digital converters (ADCs) attached to a Internet Break-Out Board (iBoB), which provides Internet protocol (IP) based communication (UDP packets) with the host computer and allows user-configurable access to the connected devices.  

\begin{figure*}[t]
\begin{center}
\includegraphics[width=5.75in]{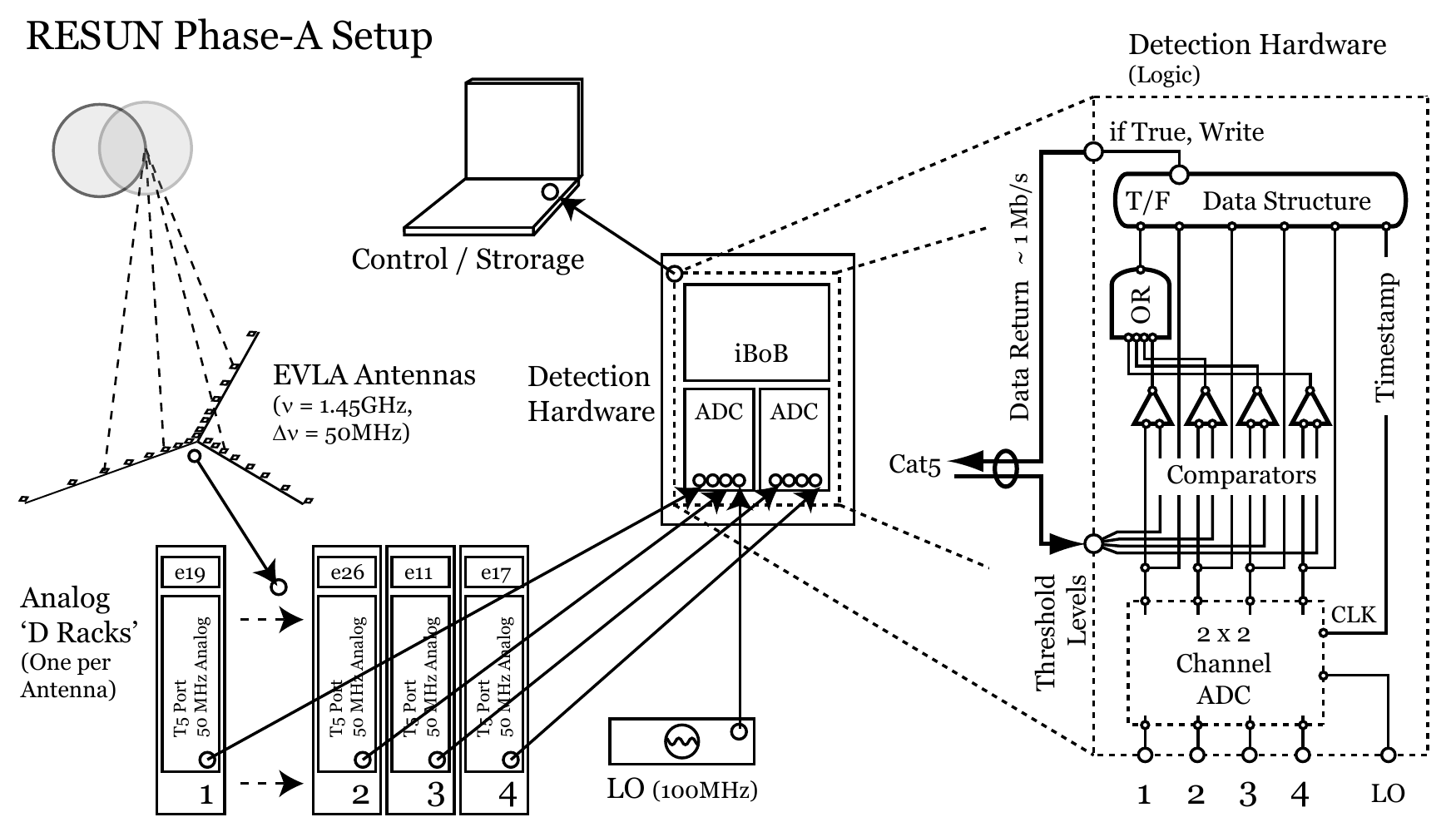}
\caption{Block diagram of RESUN pulse detection scheme for a 4-antenna sub-array.}
\label{Fig:block-diagram}
\end{center}
\end{figure*}

The raw input data stream consisted of four channels, one from each telescope, that were simultaneously sampled every 10 nanoseconds using a common sampling clock trigger.  In order to avoiding recording all bits,  we programmed the FPGA to only transfer samples (and time stamps) when the signal level in any one channel exceeded a predefined threshold level. A wave front originating from the Moon would illuminate all four antennas with differential time delays dictated by instrumental and time-dependent geometrical delays. Hence, we would record four threshold events per pulse, suitably displaced in time. A block diagram of the data acquisition system in shown in Fig. \ref{Fig:block-diagram}.

% Figure 9 - Block diagram

\begin{figure*}[h!]
\begin{center}
\includegraphics[width=5.75in]{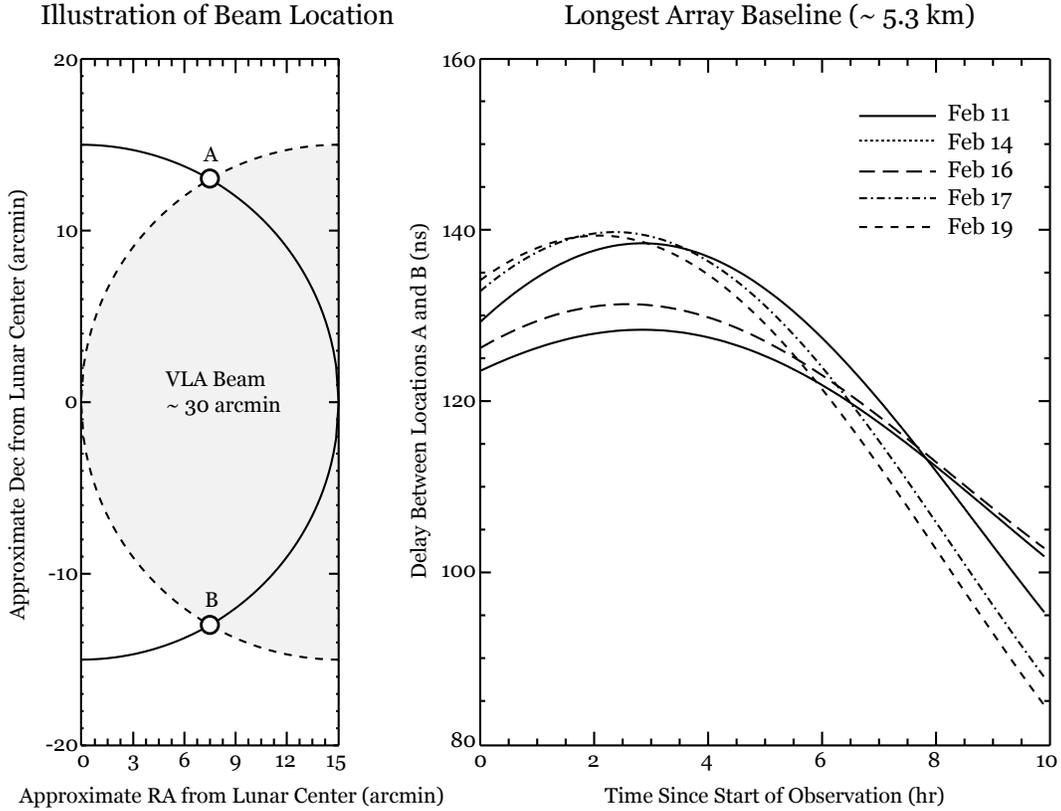}
\caption{Maximum differential delays from lunar limb sources contained within the VLA primary beam for the longest baselines in Arrays 1 and 2 at all observing epochs.}
\label{Fig:BeamDelay}
\end{center}
\end{figure*}

We searched for 4-antenna pulse coincidences by calculating the expected geometrical delays at triggered event times in a single channel. We then searched for triggered events in the other three channels using delay windows centered at the expected total (instrument plus geometrical) delays. 
We calculated the geometric delay for sources along the illuminated chord of the lunar limb using the Chapront ELP2000/82 algorithm \citep{Meeus:1998} which has  $<$ 10 arcsecond uncertainty  and $<$ 1 km uncertainty in distance. For the instrumental delays, we used the known fiber delay values. The resulting delays have an accuracy of $\sim$ 5~ns, a value smaller than the 10~ns sample time. The delay solution uncertainty is dominated instead by the delay range expected for sources distributed over the part of the lunar limb illuminated by the primary beam of the antennas (Fig. \ref{Fig:BeamDelay}). The range of delay window widths varied with projected antenna separation, but was generally in the range from 80~ns to 140~ns.  

\begin{figure*}[th!]
\begin{center}
\includegraphics[width=6.5in]{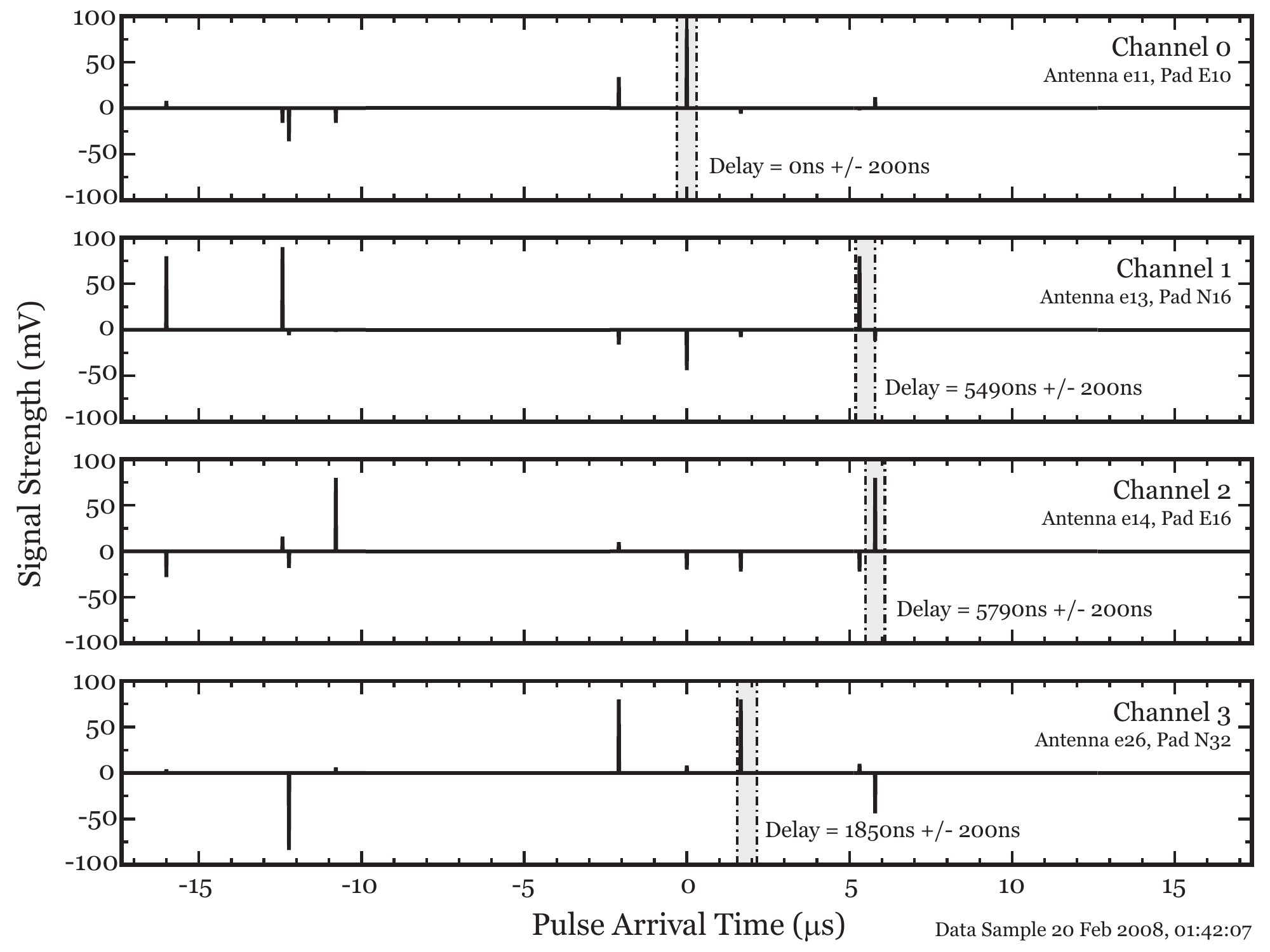}
\caption{Example of four channel accidental trigger with a wide acceptance window. The $> 3.8\sigma$ trigger signal was initiated in channel 0, while the triggers in channels 1-3 were within a $\pm200$~ns tolerance window centered on the corresponding differential (geometric plus instrumental) delays to the Moon.}
\label{Fig:FourChannel}
\end{center}
\end{figure*}

Gaussian statistics will generate many 'accidental' coincidences, as described in section \ref{sec:detection-false}, but the rate is greatly reduced as the threshold level and the number of antennas increases. For RESUN-A we chose a threshold value $\sigma_t =4.23$. With a $\pm$140 ns delay window, a four-antenna accidental coincidence would occur once every 450 observing hours, i.e., an accidental trigger probability $p\sim0.1$ in 45 hours. As a check of the pulse detection statistics, we sampled the antenna signals at a threshold value of  $\sigma_t \simeq 3.8$ and analyzed a small time range using a large delay window ($\pm200$ ns). With this delay width we expect 4-channel accidental coincidences every 10~minutes. Fig. \ref{Fig:FourChannel} shows an example of an accidental 4-antenna coincidence which disappeared when the correct delay windows were applied.

% Figure  10 - Beam delay

% Figure 11 - Four channel response

%% ANTENNA THRESHOLD, DELAY CALIBRATION TESTS
\subsubsection{Pulse detection verification tests}

Since there are no natural sources of nanosecond pulses in the sky to verify the pulse detection scheme, we built a broad-band portable pulse generator and flew it 40 m above the center of the VLA using a helium-filled balloon. The portable unit generated 25 ns wide pulses which were amplified and radiated from an omni-directional antenna. We pointed three of the subarray antennas at the pulse generator and examined the delays between pulses using both realtime oscilloscope traces and the RESUN data acquisition system. (The fourth antenna could not be tested since the balloon payload was below the minimum antenna elevation of $8\degr$). After applying the instrumental delays supplied by NRAO staff and calculating geometrical delays to the pulse generator, we verified that pulses were received at the correct delays within the expected delays to $\pm10$ ns precision.

\section{RESUN-A Observations and Results}
\label{sec:RESUN-PhaseA}

We observed the east and west limbs of the Moon using two 4-antenna subarrays for a total of 50 hours in five 10 hour sessions between 11 - 20 Feb 2008. Each sub-array used four EVLA antennas with a 50 MHz bandwidth centered at 1450 MHz.  Unfortunately, because of unexpected suppression of signals greater than $3\sigma$ for two EVLA antennas in the east sub-array (found during post-processing), this sub-array was deemed unusable and was dropped from the analysis. In addition, about five observing hours were lost as a result of bad weather, operator errors, and equipment malfunctions.

% Figure 12 - Phase A sky coverage
\begin{figure*}[h]
\begin{center}
\includegraphics[width=6.5in]{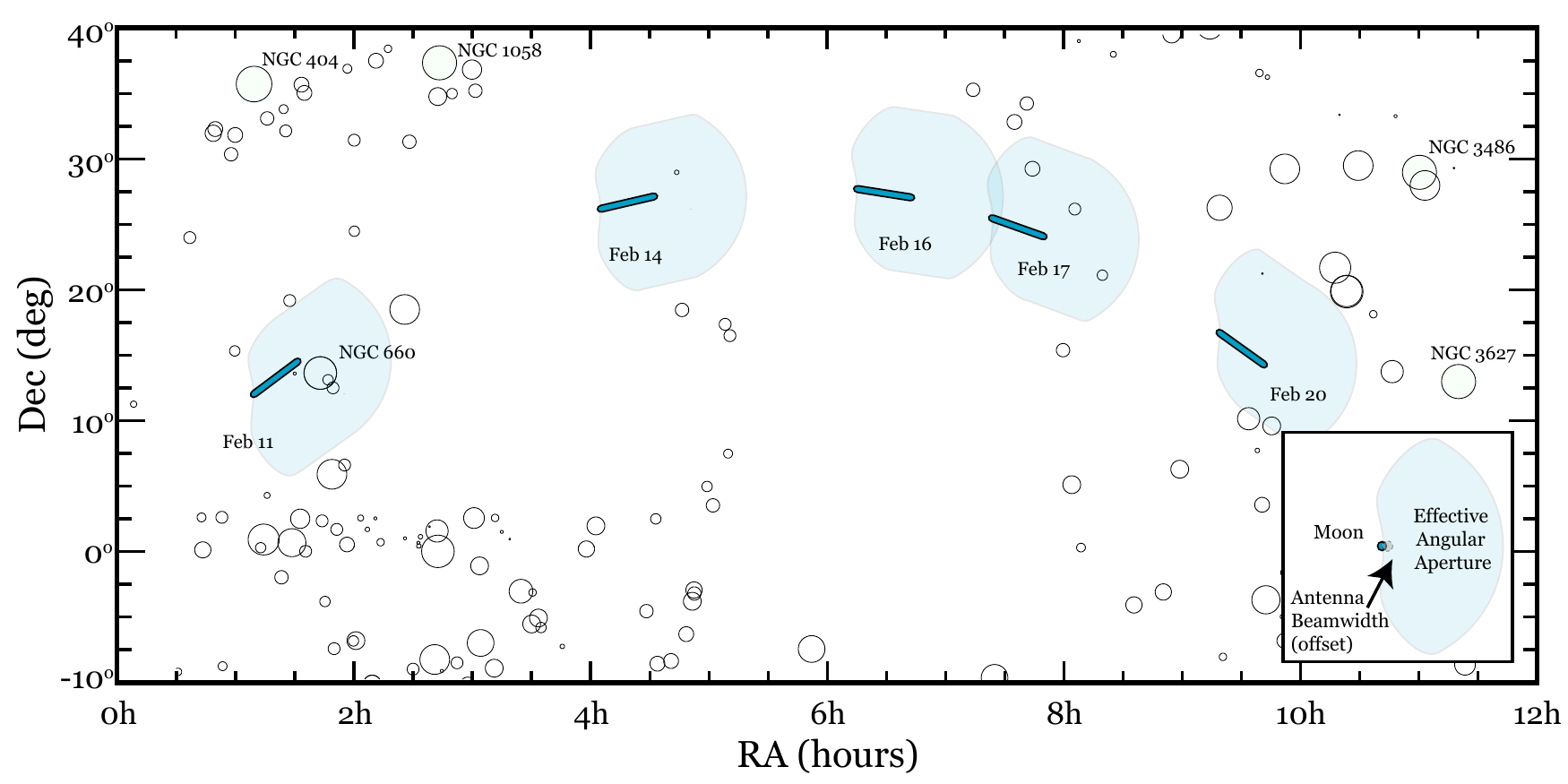}
\caption{RESUN-A sky coverage for all five observing epochs.  For each epoch, the dark green stripe is the Moon's path, while the shaded cyan region is the approximate effective aperture sky coverage.  The instantaneous aperture solid angle is shown in the inset at bottom right. Background AGN are denoted by circles,with width inversely proportional to distance.}
\label{Fig:SkyCoverage}
\end{center}
\end{figure*}

\subsection{Isotropic Flux Upper Limit}
\begin{figure}[ht!]
\begin{center}
\includegraphics[width=5.5in]{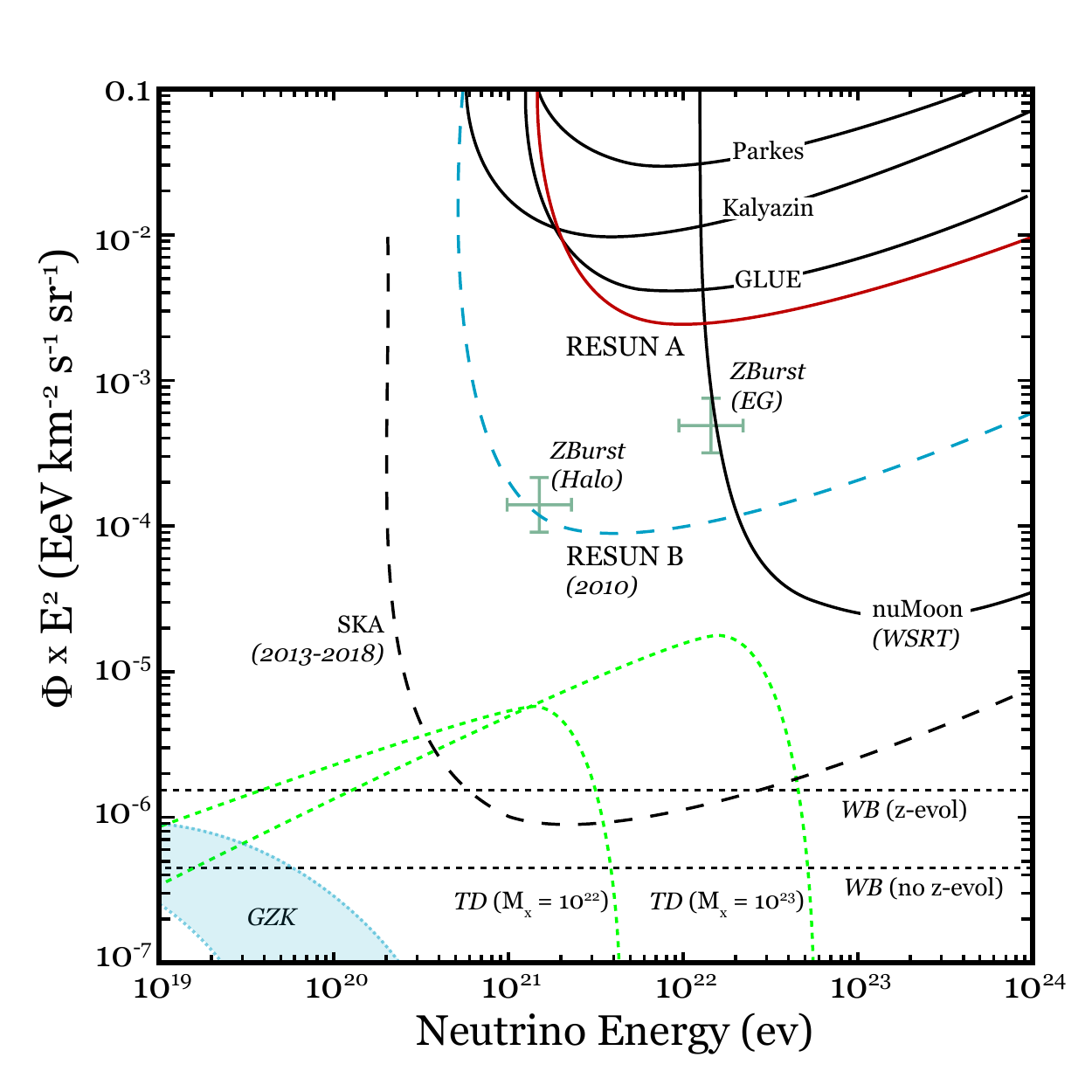}
\caption{UHE Neutrino isotropic flux upper limits determined by lunar neutrino searches: RESUN Phase -A (red line), and projected for RESUN Phase-B (blue line) along with previously reported upper limits (red dash-dot lines) from Parkes \citep{Hankins:1996,James:2007}, GLUE \citep{Gorham:2004}, Kalyazin \citep{Beresnyak:2005}, nuMoon  \citep{Scholten:2009a}. Note that all upper limits have been computed using the analytic aperture calculation of \citet{Gayley:2009} and do not necessarily agree with Monte Carlo simulations.
Also shown are cosmogenic neutrino flux predictions \citep[][and refs.]{James:2009} from Z-bursts in the galactic halo and from extragalactic background \citep{Fodor:2002a}, GZK interactions with protons and with heavy ions (blue shaded band), topological defects (TD, green short dashes), and Waxman-Bahcall upper limits for no redshift evolution and with redshift evolution (WB, black short dashes, \citet{Waxman:1999}).}
\label{Fig:UHENFlux}
\end{center}
\end{figure}

We detected no 4-station coincident events greater than a $3.98\sigma$ threshold level during an total observation period of 44.95 hours.  The corresponding upper limit to isotropic differential neutrino flux is 
\begin{equation}
E^2dN/dE < 0.003  \rm{\ EeV\ km}^{-2}\rm{s}^{-1}\rm{\ sr}^{-1}\ ,
\end{equation}
at 90\% confidence level in the neutrino energy range $21.6<$ log[E(eV)] $<22.6$. This is (marginally) lower than  the lowest published upper limits for lunar detection experiments in this energy range, and in particular confirms the GLUE upper limit reported by \citet{Gorham:2004}.  Fig. \ref{Fig:UHENFlux} shows the flux upper limit for RESUN-A vs neutrino energy, along with a number of previously reported upper limits from other lunar target searches. In addition, we show the projected flux sensitivity limits for RESUN-B  and a future SKA-class array \citep{James:2009} with 1~km$^2$ effective area at an center observing frequency 200~MHz, bandwidth 100~MHz, and 200 hours observing time.
Note that all upper limits have been calculated using the analytic effective aperture calculation described in \citet{Gayley:2009}. These may differ from upper limits reported using Monte Carlo calculations, i.e. the GLUE search \citep{Gorham:2004} and nuMoon \citep{Scholten:2009}, where the authors report upper limits nearly 10$x$ lower. This discrepancy is discussed in detail in \citet{Gayley:2009} and \citet{James:2009}.

Fig. \ref{Fig:UHENFlux} also shows expected neutrino fluxes for a variety of proposed isotropic neutrino source models. It is clear that the existing lunar searches have not yet reached sensitivity levels which could detect neutrinos from these models, but that RESUN-B will test Z-burst models \citep{Fodor:2002} and that SKA will test topological defect models and may probe the high end of the GZK model predictions, depending on total observing time and effective aperture.

% Figure 13 - UHE flux limit

% Figure 14 - Model detection
\begin{figure*}[ht!]
\begin{center}
\includegraphics[width=6.4in]{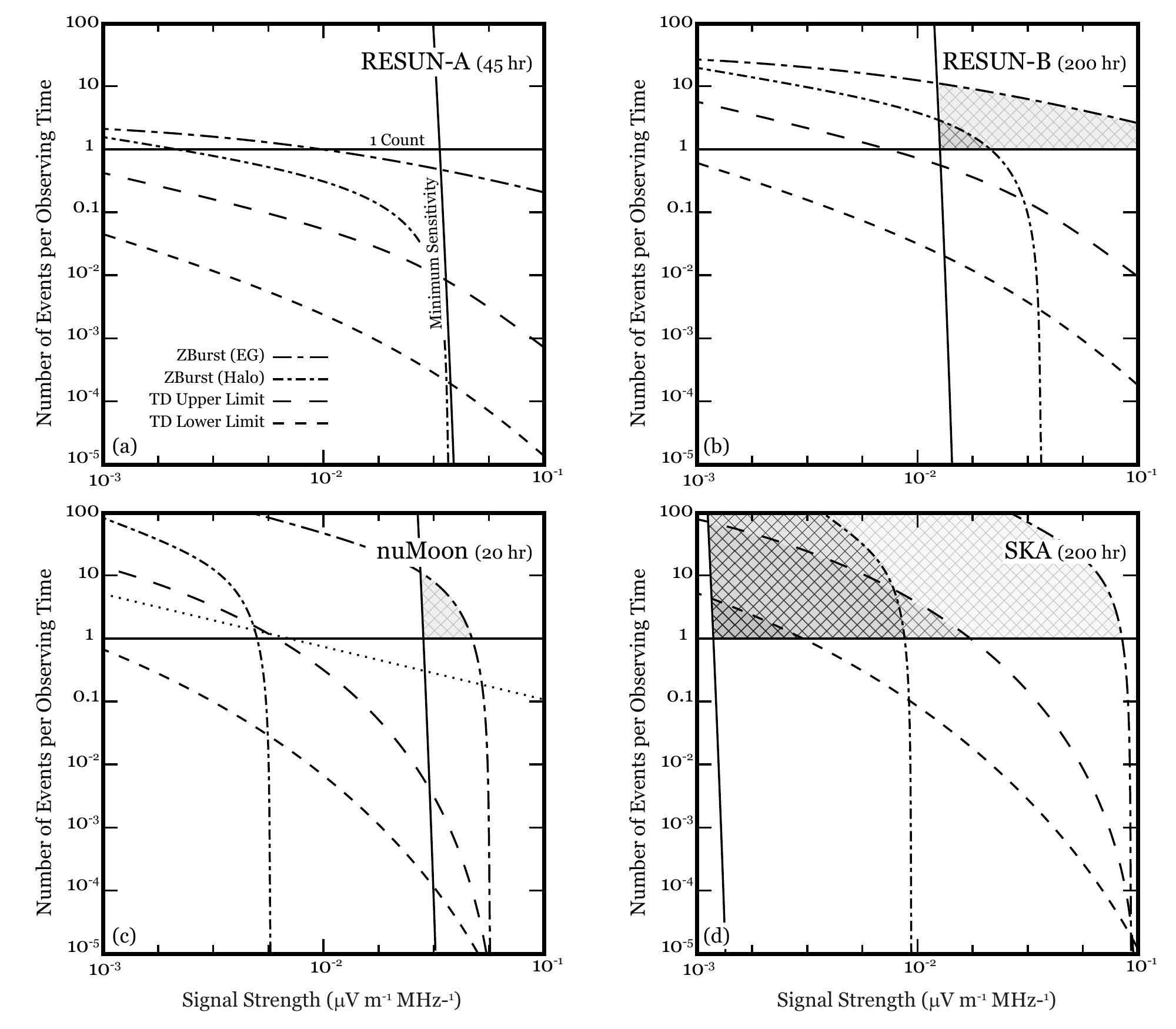}
\caption{Expected number of pulses exceeding trigger threshold versus detector sensitivity for four lunar neutrino detection experiments: (a) RESUN-A (45hr), (b) RESUN-B (200 hr), (c) nuMoon (20 hr), and (d) SKA (200 hr). In each case, the solid line is the accidental rate given by equation \ref{eqn:accid-prob}, while the labeled dotted and dashed lines are the expected rates for neutrino-production models [legend in panel (a)]. The shaded areas are regions where model counts exceed the accidental counts. In order to detect neutrinos from a given source model, the intersection between the accidental rate (nearly vertical line) and the unity count horizontal line  must lie below the model's curve.}
\label{Fig:model-detection}
\end{center}
\end{figure*}

\subsubsection{Predicted Counts vs Model plots}
Fig. \ref{Fig:model-detection} illustrates an alternative scheme to plotting isotropic flux limits versus model predictions that takes into account the predicted model-dependent detection counts in a more quantitative manner. The figure plots the number of  expected pulse detections in a fixed aperture and total observing time versus the pulse signal strength. The plot shows both the accidental rate for a given antenna array scheme and threshold (nearly vertical line), and the expected counts for a variety of source models. The intersection between the accidental rate and the horizontal one count line is a fiducial point: A search that could detect neutrinos from a given model must have its fiducial point below the model. The cross-hatched area for each of the four example searches shown in the Figure is the region for which model counts exceed accidental counts. While RESUN-A does not have any cross-hatch areas (i.e, does not test any proposed isotropic models), RESUN-B will probe both halo (dash-dot) and extragalactic (bold dash-dot) Z-burst models, with $\sim$1-10 expected counts during the 200 hour experiment. 

\subsection{Point Source Flux Upper Limit}
Although RESUN-A did not have adequate sensitivity to test isotropic neutrino source models, the non-detection result can be used to determine upper limits for UHE neutrino point source (e.g. active galaxies) fluxes along the lunar path. Fig.\ref{Fig:SkyCoverage} shows the sky coverage of RESUN-A over the entire five-day experiment, along with the sky positions of active galaxies \citep{Veron:2006}.  For point sources along the lunar path shown in the figure, the upper limit to neutrino flux is (eqn. \ref{flux-pt-upper-limit-numerical})
\begin{equation}
E^2dN/dE < 3\times10^{-4} \rm{\ EeV\ km}^{-2}\rm{s}^{-1}
\end{equation}
at 90\% confidence level in the neutrino energy range $21.6<$ log[E(eV)] $<22.6$.

\section{Summary}
\label{sec:summary}

In this paper, we present a complete, self-consistent description of the physical and technical considerations associated with searches for \^Cerenkov radio pulses from UHE neutrino interactions with the Moon.  We describe the physics of the interaction, properties of the resulting \^Cerenkov radio pulse, detection statistics, effective aperture scaling laws, and derivation of upper limits for isotropic and point source models. We report on initial results from the 
RESUN search, which uses multiple subarrays of the Expanded Very Large Array  at 1.45~GHz pointing along the lunar limb. 
We detected no lunar pulses during 44.95 observing hours. The inferred isotropic neutrino flux limit is comparable to the lowest published upper limits for lunar searches.  The full RESUN search, with 200 hours observing time and an improved data acquisition scheme, will be  be an order of magnitude more  sensitive than previous lunar-target searches, and will test Z burst models of neutrino generation.

We are grateful to Barry Clark, John Ralston, Peter Gorham, and Hallsie Reno for several illuminating discussions, Dan Mertley for technical assistance at the VLA, and the staff of CASPER for technical assistance with the data acquisition systems. The National Radio Astronomy Observatory is a facility of the National Science Foundation operated under cooperative agreement by Associated Universities, Inc.

\bibliography{RESUN-PhaseA}
\end{document}